# Signatures of charge-order correlations in transport properties of electron-doped cuprate superconductors

H. Matsuoka,[1] M. Nakano,[1,*] M. Uchida,[1] M. Kawasaki,[1,2] and Y. Iwasa[1,2]

[1] *Quantum-Phase Electronics Center and Department of Applied Physics, University of Tokyo, Tokyo 113-8656, Japan.*

[2] *RIKEN Center for Emergent Matter Science (CEMS), Wako 351-0198, Japan.*

[*] e-mail: nakano@ap.t.u-tokyo.ac.jp

**The high-temperature superconductivity in copper oxides emerges under strong influence of spin correlations in doped Mott insulators. Recent discoveries of charge-order (CO) correlations in Y-based hole-doped cuprates as well as in electron-doped cuprates suggest that charge correlations should also play an important role on the electronic states of cuprates, although those correlations have been so far detected mainly by x-ray scattering measurements. Here we show signatures of CO correlations in transport properties of electron-doped cuprates as anomalous enhancement of the metal-to-insulator crossover temperature ($T_{min}$) appears in the limited doping range near the onset of superconductivity, while it decreases exactly when superconductivity sets in. We explain this non-monotonous peak-like behavior of $T_{min}$ in terms of the evolution of the electronic states through development of CO correlations and appearance of the hole pockets in the folded Fermi surface, which impact on transport properties consecutively at different locations in the momentum space.**

# I. INTRODUCTION

Understanding the phase diagram of copper oxide superconductors has been one of the central issues of condensed-matter physics researches since the discovery of the high-temperature superconductivity [1]. Among various types of cuprate superconductors, electron-doped cuprates with the general chemical formula of $Ln_{2-x}Ce_xCuO_4$ ($Ln$ = La, Pr, Nd, Sm, and Eu) have dissimilar electronic properties to those of their hole-doped counterparts, for example $La_{2-x}Sr_xCuO_4$ (LSCO), originating from their unique crystal structures [2, 3]. Figure 1 shows the crystal structures and the electronic phase diagrams of representative hole- and electron-doped cuprates, LSCO and $Nd_{2-x}Ce_xCuO_4$ (NCCO). As compared to LSCO having octahedral coordinated copper with the $T$ structure, NCCO has square-planar coordinated copper with the so-called $T'$ structure, which is characterized by a lack of oxygen in the apical position, providing distinct properties unique to electron-doped cuprates. The most striking and well-known feature is the existence of the robust antiferromagnetic (AFM) phase in the wide doping range in their electronic phase diagrams, whereas the AFM phase in hole-doped cuprates is readily suppressed upon hole doping. Another marked difference is the stability of the charge-order (CO) correlation and its relationship to the superconducting (SC) phase. Recent resonant inelastic x-ray scattering (RIXS) measurements have revealed that the CO phase develops in NCCO with very high onset temperature ($T_{CO}$) around $T$ = 300 K, and that this newly-discovered phase coexists with the SC phase over the entire SC dome [4-6]. This is in stark contrast to hole-doped cuprates, where the CO phase exists only within the narrow doping range near $x$ ~ 0.12 with relatively low $T_{CO}$ below $T$ = 100 K under strong competition with the SC phase [7-12]. These dissimilar nature should give important insights for understanding electron-hole doping asymmetry in the high-temperature cuprate superconductors.

In this study, we examine the impact of the CO phase on transport properties of electron-doped cuprates across the insulator-to-superconductor transition. Among a variety of electron-doped cuprates with the $T'$ structure, we in particular focus on $La_{2-x}Ce_xCuO_4$ (LCCO). The electronic phase diagram of LCCO is qualitatively similar to those of well-known NCCO and $Pr_{2-x}Ce_xCuO_4$ (PCCO) due to the similarity of the crystal structure, while it shows the highest superconducting critical temperature ($T_c \sim$ 24 K) and the lowest critical doping level for superconductivity ($x_c \sim 0.08$) among all the electron-doped cuprates with the $T'$ structure [2, 3, 13]. Moreover, recent RIXS measurements have revealed the existence of the CO phase in LCCO as well, which has a similar character to that of NCCO with very high $T_{CO}$ above $T = 220$ K at $x = 0.08$ [5]. Those $T_c$ and $T_{CO}$ of LCCO are over-plotted in the phase diagram of NCCO (and PCCO) in Fig. 1(d) for comparison.

**II. EXPERIMENTAL**

We fabricated $La_{1.95}Ce_{0.05}CuO_4$ (001) epitaxial thin films on lattice matched $NdScO_3$ (110) single crystal substrates by pulsed-laser deposition [13]. For changing the doping level of LCCO, we employed an electric-double-layer transistor (EDLT) gating technique [14], which enables high-density charge accumulation up to $10^{15}$ cm$^{-2}$ on a surface of solid by application of gate voltages ($V_G$), as well as almost continuous control of the doping level in *one* sample, which should be of essential importance when exploring an electronic phase diagram and identifying relationships between different phases in a narrow doping range. We fabricated EDLTs with lightly-doped non-superconducting LCCO thin films with the Ce concentration $x = 0.05$, and examined the evolution of the electronic states by transport measurements across the insulator-to-superconductor transition while precisely shifting the doping level by changing $V_G$ (for experimental details, see Supplementary Material Section A).

**III. RESULTS AND DISCUSSION**

**A. The electronic ground state at $V_G = 0$ V**

Before focusing on the evolution of the electronic states across the insulator-to-superconductor transition, let us discuss the electronic ground state at the initial ungated state ($V_G = 0$ V). Figure 2 shows the temperature dependence of the sheet resistance ($R_s$) at $V_G = 0$ V, exhibiting the insulating ground state with the metal-to-insulator crossover temperature ($T_{min}$) of about $T = 85$ K. This insulating ground state should be attributed to the static long-range AFM phase. In this regime, large negative magnetoresistance (MR) has been known to develop below $T_{min}$ primarily due to spin correlations in the long-range AFM phase [2, 15-18]. The inset of Fig. 2 shows the variation of MR at different temperatures. There was small positive MR (less than 0.1 %) observed at higher temperature above $T > 100$ K, while large negative MR detected below $T = 60$ K. The magnitude of this negative MR became larger with lowering temperature, suggesting that the ground state should have a magnetic order. Moreover, the onset temperature of this negative MR ($T_{MR}$) coincides with $T_{min}$, indicating that $T_{min}$ at this low-doping regime should correspond to the onset of a magnetic order in the static long-range AFM phase. We note that $T_{MR}$ at the initial ungated state is lower than the Néel temperature ($T_N$) of NCCO, which is around $T = 220$ K at $x = 0.05$, but close to $T_N$ of LCCO near the similar doping level. The magnetic orders in LCCO have been less studied compared to those in NCCO and PCCO mainly due to the fact that LCCO is stable only in thin-film form, which is not suitable for evaluation of the magnetic structure by neutron scattering measurements. However, there are a few reports discussing magnetic orders in this compound, either by in-plane angular magnetoresistance (AMR) [16] or by low-energy muon spin rotation (LE-μSR) [19]. Those studies show that $T_N$ of LCCO is around $T = 110$ K at $x = 0.07$, which is close to $T_{MR}$ obtained for our sample at the ungated state. We also note that the AMR and the LE-μSR results suggest that there are two types of the AFM orders in LCCO, either long-range or short-range as is the case in NCCO [20], which will be discussed in more details later in Section III-E.

**B. Gate-induced insulator-to-superconductor transition**

Let us now discuss the gating effects on the transport properties. Figure 3 shows the temperature dependence of $R_s$ at different $V_G$, demonstrating the gate-induced insulator-to-superconductor transition in LCCO. The $R_s$-$T$ curves showed negligible variation in the small $V_G$ region, whereas $R_s$ began to decrease above the threshold voltage of about $V_G = 1.5$ V. The superconductivity emerged at $V_G = 2.60$ V and $T_c$ increased up to 22 K, which is close to the maximum $T_c$ (~ 24 K) of chemically-doped LCCO films with the Ce concentration $x = 0.11$ [13]. The observed gating effects were highly reversible (see the inset of Fig. 3) and reproducible (see Supplementary Material Section B), enabling detailed examination of electronic properties on *one* sample without changing any other parameters.

**C. The electronic phase diagram**

Based on the above results, we constructed an electronic phase diagram as shown in Fig. 4(a). Also shown are the $V_G$ dependences of $T_c$ and $T_{min}$. The resulting phase diagram indicates that the ground state could be continuously evolved from insulating to SC by increasing $V_G$. Furthermore, we found that $T_{min}$ exhibits anomalous peak-like behavior with increasing $V_G$, whose peak is located exactly at the onset of superconductivity. Given that the insulating ground state at the low-doping regime should be the static long-range AFM phase as discussed in Section III-A, and that $T_{MR}$ was decreased with increasing the doping level as will be discussed in Section III-F, this non-monotonous behavior of $T_{min}$ suggests that there should be another insulating ground state setting in near the boundary between the long-range AFM and the SC phases. Hereafter, we discuss possible origins of this peak behavior of $T_{min}$ by separately focusing on the increasing and the decreasing regimes.

**D. Increasing $T_{min}$ regime** For the increasing regime, we propose that this should arise from the development of CO correlations, which have been recently discovered and deeply investigated in NCCO bulk single crystals [4-6]. The important findings there are, (1) the CO is formed between the parallel segments of the small electron pockets near $k \sim (\pi, 0)$ in the Brillouin zone, and (2) the CO sets in at around $x = 0.05$, sharply grows up with very high onset temperatures ($T_{CO}$) as the doping level $x$ increases, and survives up to $x = 0.17$ while keeping such high $T_{CO}$ [Fig. 1(d)]. Considering that the formation of the CO near $k \sim (\pi, 0)$ should impact carrier transport through reduction of the density of states at the Fermi level and/or the increase of the carrier scattering rate due to charge fluctuations, a crossover from metallic to insulating (or less metallic) behavior is expected to emerge upon cooling below $T_{CO}$. Given that the CO phase in LCCO has a similar character to that of NCCO with comparably high $T_{CO} > 220$ K [5], and that $T_{CO}$ in NCCO increases with increasing the doping level in this regime, we conclude that the increase of $T_{min}$ in the present study should be attributed to the development of the CO phase in LCCO. We note that $T_{min}$ characterized by the transport measurements in this study was far below $T_{CO}$ defined by the RIXS measurements in the previous studies. This large discrepancy between the transport and the diffraction results has been also observed for CO correlations in Y-based hole-doped cuprates [7-9, 21, 22], where $T_{CO}$ from the diffraction measurements was attributed to the onset of the CO fluctuations above the SC dome, while the characteristic temperature evaluated by the transport measurements was assigned to the onset of the static CO orders below the SC dome. The obtained discrepancy in the present study therefore implies that there might be two types of CO correlations in electron-doped cuprates as well, either fluctuations or static orders, and $T_{CO}$ corresponds to the onset of the CO fluctuations while $T_{min}$ reflects that of the static CO orders, although further investigations are needed to support this interpretation.

**E. Decreasing $T_{min}$ regime**

The CO phase should persist up to higher doping level according to the previous diffraction experiments, whereas the increase of $T_{min}$ characterized by the transport measurements should end when the system enters the SC regime due to development of the hole pockets at $k \sim (\pi/2, \pi/2)$ in the Brillouin zone, which is also a unique and general feature of electron-doped cuprates widely observed in NCCO [23, 24], PCCO [25, 26], and LCCO [27]. Figure 4(b) shows the variation of the Hall coefficients ($R_H$) of the same device measured at the normal states ($T = 30$ K) as a function of $V_G$, indicating (I) the lightly-doped negative $R_H$ regime, (II) the intermediately-doped regime showing the increase of $R_H$, and (III) the heavily-doped positive $R_H$ regime. Considering the correspondence between the variation of $R_H$ and the evolution of the Fermi surface evaluated by angle-resolved photoemission spectroscopy measurements [23-27], we argue that this non-monotonous change in $R_H$ originates from the evolution of the Fermi surface from that with the small electron pockets located at $k \sim (\pi, 0)$ under strong influence of the long-range AFM order [regime (I)] to the large hole-like cylindrical Fermi surface centered at $k \sim (\pi, \pi)$ above the AFM quantum critical point (QCP) where the AFM order and the resulting band folding no longer exist [regime (III)]. In the intermediate regime (II), $R_H$ is increased from negative to positive, which is characterized by the appearance of the hole pockets (or 'Fermi arcs') at $k \sim (\pi/2, \pi/2)$ and their development toward $k \sim (\pi, 0)$. This development of the hole pockets should recover the density of states at the Fermi level independently from the formation of the CO near $k \sim (\pi, 0)$, leading to the decrease of $T_{min}$. The $V_G$ dependences of $T_{min}$ and $R_H$ plotted in Fig. 4(a) and 4(b) show that the decrease of $T_{min}$ occurs when $R_H$ begins to increase, in other words, when the system enters the regime (II), verifying that the development of the hole pockets should be responsible for the decrease of $T_{min}$. The most plausible origin that determines $T_{min}$ in this regime (II) should be the short-range AFM order [black dashed line in Fig. 1(d)], which is known to commonly exist in electron-doped cuprates including NCCO [20], PCCO [15, 28], and LCCO [16, 19, 27] over the SC dome and terminate at the AFM QCP within the SC dome. The observed coincidence between the decrease of $T_{min}$ and the increase of $T_c$ therefore reveals inseparable relationship between AFM spin correlations and superconductivity in electron-doped cuprates.

### F. Evolution of $T_{MR}$

We note that the increase of $T_{min}$ discussed in Section III-D should be less relevant to the evolution of the long-range AFM phase with increasing the doping level. Figure 5(a) shows the variation of MR at different $V_G$ at $T = 30$ K, demonstrating a crossover from negative MR to positive MR with increasing the doping level. Given that this crossover temperature should correspond to $T_{MR}$ as we discussed in Section III-A, $T_{MR}$ should be higher than 30 K below $V_G = 2.76$ V, whereas $T_{MR}$ should be lower than 30 K above $V_G$ = 2.78 V. Considering that $T_{min}$ is much higher than 30 K around this doping level [Fig. 5(b)], we conclude that the observed enhancement of $T_{min}$ should not originate from development of the long-range AFM phase but from the onset of the CO phase as we discussed in Section III-D.

**G. Generality of peak-like behavior of $T_{min}$ in electron-doped cuprates**

Taken above interpretations together, we believe that a predominant effect that determines $T_{min}$ evolves from spin correlations in the static long-range AFM phase at the initial ungated states (Section III-A) to charge correlations in the CO phase with increasing the doping level (Section III-D), while the short-range AFM order governs the decrease of $T_{min}$ at the higher doping level (Section III-E) through development of the hole pockets on the Fermi surface [15, 16, 20, 27, 28]. This non-monotonous variation of $T_{min}$ has been widely detected but not fully addressed in the previous studies on NCCO [29] and PCCO [26, 30]. Figures 6(a) and 6(b) summarize the variation of $T_{min}$ and $T_c$ for electron-doped cuprates (PCCO and NCCO) while changing the doping level either by chemical doping [26, 29] or by electrostatic doping using EDLT [30], respectively. A peak-like behavior of $T_{min}$ similar to our observation on LCCO-EDLT was clearly shown near the onset of superconductivity irrespective of the material (PCCO or NCCO) and the doping process (chemically or electrically), suggesting that this is an intrinsic and general feature of electron-doped cuprates. Moreover, very interestingly, the EDLT study on monolayer PCCO presented dramatic enhancement of $T_{min}$ [30], indicating strong dimensionality effect of CO correlations probably due to enhanced fluctuations with reduced dimensions. In contrast, as shown in Figs 6(c) and 6(d), $T_{min}$ in hole-doped cuprates decreased monotonically with increasing the doping level both for chemically- and electrically-doped LSCO [31, 32], suggesting remarkable difference between hole- and electron-doped cuprates. The origin of this dichotomy might be related to distinct temperature scales of the CO phase and its doping dependence.

**IV. SUMMARY**

Our results indicate development of CO correlations in the underdoped regime of electron-doped cuprates near the onset of superconductivity, which should play an essential role on the electronic states and the emergence of superconductivity in those compounds. We emphasize that it is the ion-gating technique enabling almost continuous tuning of the doping level that makes such detailed examination of the phase diagram possible, demonstrating the usefulness of this method for exploration of electronic states of condensed matters. Further investigations of the electronic phase diagrams by combination with optical measurements that characterize symmetry breakings should unveil important aspects for the origin of the high-temperature superconductivity in cuprate superconductors, which has been so far hindered by inherent difficulties in preparation of various samples having different doping levels with minimal variations in sample quality and stoichiometry.

**ACKNOWLEDGEMENTS**

This work was partly supported by Grants-in-Aid for Scientific Research (Grant Nos. 25000003 and 15H05499) from the Japan Society for the Promotion of Science (JSPS). We are grateful to K. S. Takahashi, D. Maryenko, M. Nakamura, J. Matsuno, and Y. Kozuka for help with thin film growth. H.M. was supported by JSPS through Program for Leading Graduate Schools (MERIT).

**Figure captions**

FIG. 1. (a) Crystal structure of hole-doped cuprate superconductor, $La_{2-x}Sr_xCuO_4$ (LSCO), with octahedral coordinated copper (*T*-structure). (b) Electronic phase diagram of hole-doped cuprates with *T*-structure, LSCO and Eu-doped LSCO. Black solid line is the Néel temperature ($T_N$) of LSCO determined from magnetic susceptibility measurements [31], while blue circles denote the superconducting critical temperature ($T_c$) of LSCO [31]. Green circles are the pseudogap onset temperature ($T^*$) of LSCO (filled) and Eu-doped LSCO (open) determined from the upturn in the Nernst coefficient [33]. Red circles are the onset temperature of charge order ($T_{CO}$) of LSCO (filled) and Eu-doped LSCO (open) determined from x-ray scattering measurements [10, 12] (c) Crystal structure of electron-doped cuprate superconductor, $Nd_{2-x}Ce_xCuO_4$ (NCCO), with square-planar coordinated copper (*T'*-structure). (d) Electronic phase diagram of electron-doped cuprates with *T'*-structure, NCCO and $Pr_{2-x}Ce_xCuO_4$ (PCCO). Black solid and dashed lines are long- and short-range AFM orders of NCCO, respectively, determined from inelastic magnetic neutron-scattering measurements [20]. Blue circles are $T_c$ of NCCO [5]. Green circles are $T^*$ of NCCO (filled) and PCCO (open) determined from optical conductivity measurements [29, 34]. Red circles are $T_{CO}$ of NCCO determined from x-ray scattering measurements [5]. $T_c$ and $T_{CO}$ of $La_{2-x}Ce_xCuO_4$ (LCCO) are also shown as blue and red stars [5, 13], respectively.

FIG. 2. The sheet resistance ($R_s$) of LCCO thin film (black, left) and magnetoresistance (MR) defined as $[R(B)-R_0]/R_0$ at $B = 9$ T (red, right) with the field perpendicular to the plane as a function of temperature at the initial ungated state ($V_G = 0$ V). The inset shows MR at different temperatures.

FIG. 3. $R_s$ of LCCO thin film as a function of temperature with different gate voltages ($V_G$) applied through the ionic liquid layer, demonstrating the gate-induced insulator-to-superconductor transition. The inset shows the $R_s$-$T$ curves taken at $V_G = 0$ V before and after the gating experiments, verifying good reversibility of the device.

FIG. 4. (a) The electronic phase diagram of LCCO constructed by the data obtained in this study, where $R_s$ is plotted in $T$-$V_G$ plane. Blue circles are $T_c$ defined as the temperature at which $R_s$ drops by 90 % to its normal-state value. Black circles are the metal-to-insulator crossover temperature ($T_{min}$) defined as the temperature at $dR_s/dT = 0$. (b) The variation of the Hall coefficients ($R_H$) as a function of $V_G$ measured at normal states, $T = 30$ K.

FIG. 5. (a) MR with the field perpendicular to the plane at different $V_G$ measured at normal states, $T = 30$ K. (b) The onset temperature of negative MR ($T_{MR}$) both at the ungated and the gated states (red circles) together with the variation of $T_{min}$ (black circles) and $T_c$ (blue circles) with increasing the doping level.

FIG. 6. (a) $T_{min}$ (red circles) and $T_c$ (black circles) of PCCO thin films [26] (filled circles) and NCCO bulk single crystals ($T_{min}$ from ref. [29] and $T_c$ from ref. [5]) (open circles) as a function of the doping level $x$. (b) Those of PCCO-EDLT [30] as a function of the doping level $V_G$. (c) $T_{min}$ (blue circles) and $T_c$ (black circles) of LSCO bulk single crystals [31] as a function of $x$. (d) Those of LSCO-EDLT [32] as a function of $V_G$.

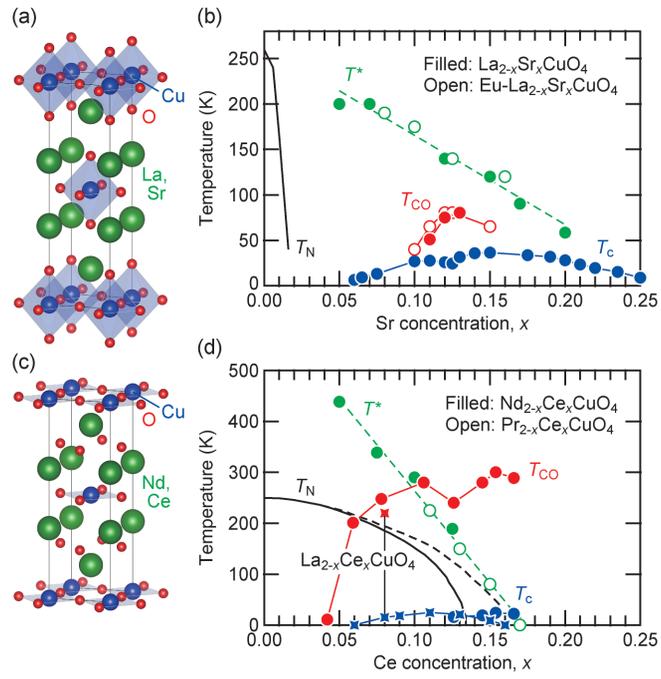

Fig. 1 H. Matsuoka et al.

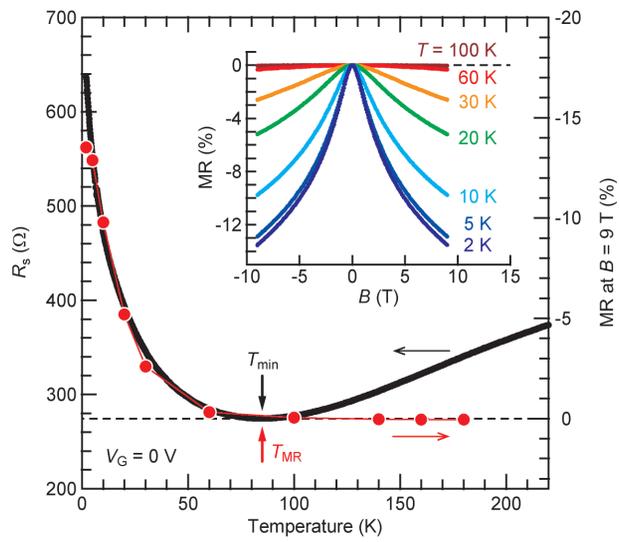

Fig. 2  H. Matsuoka et al.

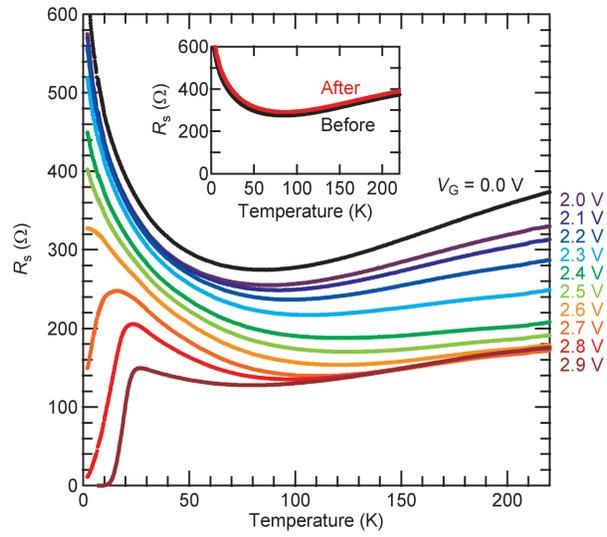

Fig. 3  H. Matsuoka et al.

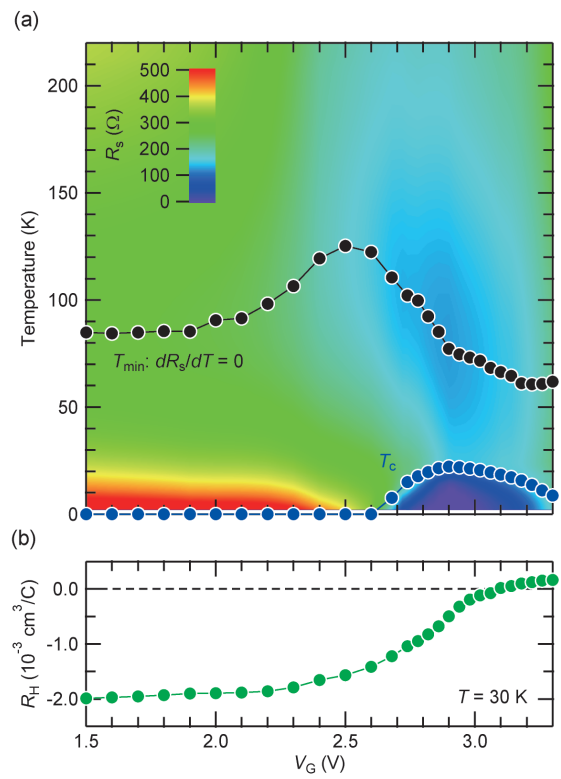

Fig. 4  H. Matsuoka et al.

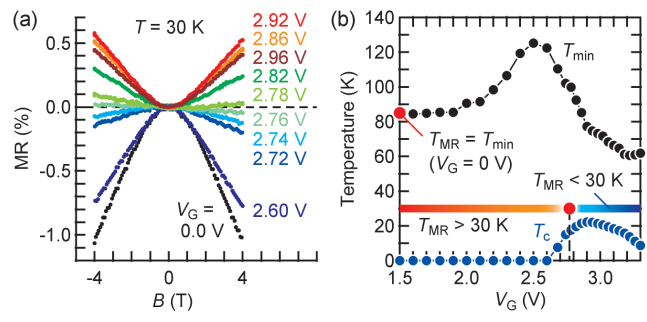

Fig. 5 H. Matsuoka et al.

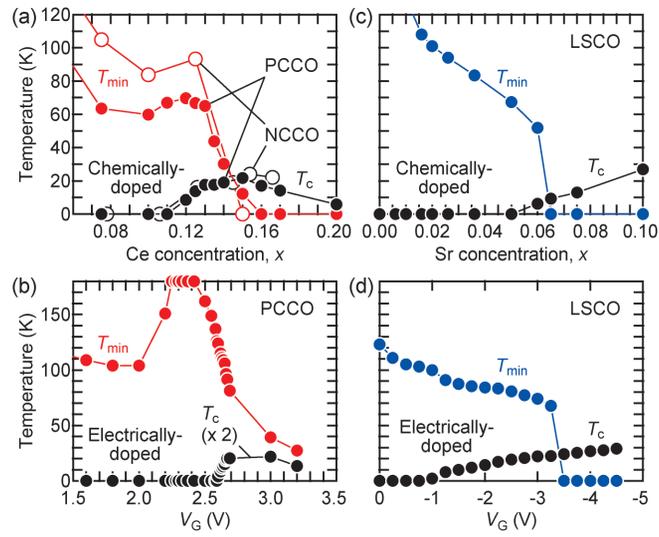

Fig. 6  H. Matsuoka et al.